\documentclass[]{aa} 
\usepackage{graphicx}
\usepackage{txfonts}
\usepackage{natbib}
\begin{document}
%
\title{Deriving the radial-velocity variations induced by stellar activity from high-precision photometry }
\subtitle{Test on HD~189733 with simultaneous MOST/SOPHIE data}
  
\authorrunning{A. F. Lanza et al.}
\titlerunning{Synthesizing the activity-induced RV variations of HD~189733}

   \author{A.~F.~Lanza\inst{1} \and I.~Boisse \inst{2,3} \and F. Bouchy\inst{3,4} \and A.~S.~Bonomo\inst{1,5} \and C.~Moutou\inst{5}
          }

   \institute{INAF-Osservatorio Astrofisico di Catania, Via S.~Sofia, 78, 95123 Catania, Italy\\
              \email{nuccio.lanza@oact.inaf.it}
\and
Centro de Astrof\'isica da Universidade do Porto, Rua das Estrelas, 4150-762 Porto, Portugal 
\and
Institut d'Astrophysique de Paris, Universit\'e Pierre et Marie Curie, UMR7095 CNRS, 98bis bd. Arago, 75014 Paris, France
\and
Observatoire de Haute Provence, CNRS/OAMP, 04870 St Michel l'Observatoire, France
\and 
Laboratoire d'Astrophysique de Marseille (UMR 6110),
Technopole de Ch\^{a}teau-Gombert,
38 rue Fr\'ed\'eric Joliot-Curie,
13388 Marseille cedex 13, France     
    }

   \date{Received ; accepted }

 
  \abstract
   {Stellar activity induces apparent radial velocity (RV) variations in late-type main-sequence stars that may hamper the detection of low-mass  planets and the measurement of their mass.}
   {We use simultaneous measurements of the active planet host star HD~189733 with high-precision optical photometry by the MOST satellite and high-resolution spectra by SOPHIE. We apply on this unique dataset a spot model to predict the activity-induced RV variations and compare them with the observed ones.}
   {The model is based on the rotational modulation of the stellar flux. A maximum entropy regularization is applied to find a unique and stable solution for the distribution of the active regions versus stellar longitude. The RV variations are synthesized considering the effects on the line profiles of the brightness perturbations due to dark spots and bright faculae and the reduction of the convective blueshifts in the active regions. }
   {The synthesized RV time series shows a remarkably good agreement with the observed one although variations on timescales shorter than $4-5$ days cannot be reproduced by our model. Persistent active longitudes are revealed by the spot modelling. They rotate with slightly different periods yielding a minimum relative amplitude of the differential rotation of $\Delta \Omega / \Omega = 0.23 \pm 0.10 $. Moreover, several active regions with an evolution timescale of $2-5$ days and an area of $0.1-0.3$ percent of the stellar disc are detected. } 
  {The method proves capable of reducing the power of the activity-induced RV variations by a factor from 2 to 10 at the rotation frequency and its harmonics up to the third. 
  Thanks to the high-precision space-borne photometry delivered by CoRoT, Kepler, or later PLATO, it is possible to map the longitudinal distribution of active regions in late-type stars and apply the method presented in this paper to reduce remarkably the impact of stellar activity on their RV jitter allowing us to confirm the detection of low-mass planets or refine the measurement of their mass.}


   \maketitle
%

\section{Introduction}


Several studies have recently addressed the perturbation of the RV induced by stellar activity.  This jitter may severely hamper the detection of low-mass planets or the measurement of the mass of those found by the method of the transits.  \citet{Lagrangeetal10}, \citet{Meunieretal10a}, and \citet{Meunieretal10b} have considered the Sun as a star finding that the reduction of convective blueshifts in active regions, due to the quenching of turbulent motions by the   photospheric magnetic fields, has a remarkable effect inducing variations with an amplitude up to $5-10$~m~s$^{-1}$. 
\citet{Dumusqueetal11} have extended those studies introducing a model for the impact of dark spots on the RV variations of solar-like stars based on the statistics of sunspot groups. 

Several techniques have been proposed to reduce the impact of the RV perturbation induced by 
stellar activity. They can be classified into two broad groups. Those in the first group exploit the correlation between the RV perturbation and other properties of the cross-correlation function used to measure the RV itself, such as the mean line bisector shift and/or its FWHM \citep[e.g., ][]{Meloetal07,Boisseetal09,Pontetal11}. The main limitation of such techniques is the lack of a tight correlation between the RV perturbation and those proxies for the average spectral line distortion, as demonstrated by, e.g., \citet{Boisseetal11}. The other group consists of the techniques that exploit the modulation of the activity-induced perturbation by the stellar rotation. Most of the power due to the activity-induced perturbation is indeed concentrated at the rotation frequencies and its first two or three harmonics. Therefore,  it is possible to filter it out leaving the signal due to the reflex orbital motion induced by a planetary companion, unless its orbital period falls close to such frequencies, and to fit simultaneously activity and planetary signals \citep[see ][ for details]{Boisseetal11}. This approach has been recently applied, among others, to detect a second non-transiting planet around the active star \object{CoRoT-7} and to refine the measurement of the mass of its transiting telluric planet CoRoT-7b \citep[see ][]{Quelozetal09,Boisseetal11}. 

In addition to those methods, another approach to reduce the impact of stellar activity can be based on the models  of the active region distribution derived from  high-precision photometry such as that provided by the space-borne telescopes of MOST, CoRoT, or Kepler for their target stars. From their light curves, it is possible to derive the longitudes of the active regions and their area, tracing their evolution along successive rotations. This information can be used to model  the RV perturbation induced by stellar activity. \citet{Lanzaetal10} have applied this approach to the  light curve of CoRoT-7 to predict the power spectrum of the RV perturbation induced by stellar activity and derive the significance of the signals attributed to the telluric planets CoRoT-7b and CoRoT-7c. The main limitation of that study was the lack of simultaneity between the CoRoT light curve and the HARPS RV time series analysed to extract the reflex motions induced by the two planets.

 Here we further explore this approach by means of simultaneous photometric and RV measurements of HD~189733 making use of the MOST light curve and the SOPHIE RV time series published by \citet{Boisseetal09}. They span about one month in the summer of 2007, corresponding to two full rotations of the star with a period of $P_{\rm rot} \sim 12$~days \citep{HenryWinn08}. \object{HD~189733} is accompanied by a Jupiter-mass transiting planet with an orbital period of 2.218 days. With an effective temperature of $\sim 5000$~K { and an average chromospheric index $S_{\rm HK} = 0.525$}, this K1 main-sequence star is remarkably active \citep{Bouchyetal05,Moutouetal07,Boisseetal09}. Its radial velocity (hereafter RV) measurements are then perturbed with variations up to $\sim 20-30$~m~s$^{-1}$, as evident in the residuals of the RV  best fit of the planetary orbit \citep{Boisseetal09}. 

Our spot modelling technique allows us to retrieve the location and the evolution of the photospheric active regions of HD~189733 finding also evidence of a remarkable stellar differential rotation, as recently found by \citet{Faresetal10} by means of Doppler Imaging techniques. The information on the starspot longitude and evolution is used to synthesize the RV variations that can be compared with the simultaneous SOPHIE measurements to assess the advantages and limitations of the technique. We find that the modulation of the RV perturbation induced by stellar activity is reproduced remarkably well on timescales comparable with the stellar rotation period but it is not possible to model the variations with a timescale of $1-2$ days, i.e., much shorter than the stellar rotation period. This is an unavoidable consequence of the fact that the rotational modulation of the optical flux is used to map starspots. On the other hand, the main advantage of this method is the remarkable reduction of the perturbations at frequencies comparable with the stellar rotation frequency and its low-order harmonics, that is an important improvement over previously proposed approaches.

The next section describes the observational dataset. Then, we define our model: 1) how active regions locations are derived from lightcurve; 2) how we compute the distortion of the line profiles due to the brightness perturbations owing to dark spots and bright faculae and the reduction of the convective blueshifts in the active regions; and 3) how the distributions of active regions derived from the spot model fitted to the light curve are used to synthesize RV variations. The fourth section gives the parameters chosen for our modelling of HD~189733. The results, given in  Section 5, are discussed in the final section.

\section{Observations}
\label{observations}

\subsection{MOST photometry}
\label{photometry}

 In 2007, HD~189733 was monitored by the MOST  (Microvariability and Oscillations of STars) microsatellite for 30.4994 days covering $\sim$~2.5 rotation cycles of the star  from  HJD~2454298.55874 to 2454329.05818.  
 MOST  was launched in 2003 and was designed to observe nearby, relative bright stars through a single broadband (350-700~nm) filter using a 15/17.3-cm Rumak-Matsutov telescope \citep{Walkeretal03,Matthewsetal04}. 
 The photometry was collected in the so-called MOST's Direct Imaging mode, where a defocussed stellar image was recorded on a subraster of a single  CCD.  The data were reduced as described in \citet{Roweetal06}. The stellar fluxes were extracted by aperture photometry and the raw instrumental light curve was decorrelated against the sky background and also against the location of the centroid of the point-spread function on the CCD. 
The satellite had an 820~km altitude Sun-synchronous polar orbit of 101.4 minutes and the data were binned to this period. The transits of the known planet were removed from the MOST lightcurve. This led to  a time series consisting of 414 points.  As seen in the top panel of Fig.~\ref{lc_bestfit}, HD~189733 exhibited in 2007 a $\sim$~1\% variation in flux, which is much less than the $\sim$~3\% variation observed in 2006 by the same telescope  \citep{Miller-Riccietal08,Crolletal07}. Moreover, a decreasing long-term trend was seen that was not exhibited by the comparison stars, suggesting that the drift was real. 

 We assume the maximum observed flux at HJD~2454300.31968 as the unit of measure for the flux and normalize the light curve to that value. The mean error of  the binned points is $1.15 \times 10^{-4}$ in relative flux units. However, since MOST photometry is non-differential and can experience drifts on daily or weekly timescales, such a value should be considered as a lower limit to the photometric precision and does not take into account systematic errors. For example, \citet{Crolletal07} pointed out that the stray light background can  be modulated at a period of 1 day and its first harmonic (due to the Sun-synchronous orbit).
The flux level corresponding to the star without spots is unknown, therefore we assume the maximum observed flux as the reference level for the spot modelling, i.e., we assume that it corresponds to the flux of the unspotted star in the MOST passband. 
We can also mention that the M-dwarf companion of HD~189733 \citep{Bakosetal06} does not contaminate appreciably the photometric signal.

\subsection{SOPHIE radial velocity measurements}
\label{rv_measurements}

 SOPHIE is a high-resolution spectrograph mounted on the  1.93-m telescope at the Observatoire de Haute-Provence, France \citep{Perruchotetal08,Bouchyetal09b}. 
The SOPHIE observations of HD~189733 were conducted simultaneously with the MOST photometry. 
Usually two exposures per night were gathered during more than one month (from 12 July to 23 August 2007) to sample  both the short orbital period ($\sim 2.2$~days) and the longer rotation period of the star ($\sim 12$~days). 
The observations were conducted in the high-resolution mode ($\lambda / \Delta \lambda \sim 75000$) and with an average signal-to-noise ratio (SNR) per pixel of about 80 at $\lambda =550$~nm.  The journal of the observations and the instrument setup were reported in Sect.~2 of \citet{Boisseetal09}.

For the present work, the original spectra were reduced again with a new pipeline that, among others, corrects for the effects of the charge transfer inefficiency of the CCDs \citep{Bouchyetal09a} and removes part of the spectrum that  may contain telluric lines before applying the correlation mask. This gives slightly different weights to the red and blue parts of the spectrum for the determination of the RV through the standard cross-correlation method, resulting in a RV offset of +21.6~m~s$^{-1}$ with respect to the RV values reported in \citet{Boisseetal09}. The recently reported \textit{seeing effect} \citep{Boisseetal10} was not well constrained and observed in these data, therefore it is not corrected. { An average error of 3.9~m~s$^{-1}$ for the RV measurements is then a good estimate of the accuracy of our values accounting for the photon-noise error, the uncertainty in the wavelength calibration, and SOPHIE systematics. The effects of  the systematics is taken to be 3~m~s$^{-1}$ for all the measurements, derived from the standard deviation of the RV of the most stable stars observed with SOPHIE,  
some of which have been observed with HARPS to have a stability better than 1~m~s$^{-1}$.
 These stars have more than 10 measurements and have been observed for more than six months in order to take into account the long-term systematics. We note that  new fiber scramblers have just been installed on SOPHIE (June 2011), and preliminary tests showed a significant improvement in the RV accuracy.  }
To compute the RV perturbations due to stellar activity, the orbital solution has been subtracted. The adopted orbital elements are those given by \citet{Boisseetal09}.

\section{Light curve and RV modelling}

\subsection{Spot modelling of wide-band light curves}
\label{spotmodel}

The reconstruction of the surface brightness distribution from the rotational modulation of the stellar flux is an ill-posed problem, because the variation of the flux vs. rotational phase contains information only on the distribution of the brightness inhomogeneities vs. longitude. The integration over the stellar disc effectively cancels any latitudinal information, particularly when the inclination of the rotation axis along the line of sight is close to $90^{\circ}$, as in the present case \citep[see Sect.~\ref{model_param} and ][]{Lanzaetal09a}. Therefore, we need to include a priori information in the light curve inversion process to obtain a unique and stable map. This is  done by computing a maximum entropy (hereinafter ME) map, which has been proven to  best reproduce active region distribution and area variations in the case of the Sun \citep[cf. ][]{Lanzaetal07}. For a different modelling approach, based on discrete starspots, see, e.g., \citet{Mosseretal09}. 

In the present model, the stellar surface is subdivided into  elements, i.e., in  200  squares of side $18^{\circ}$, with  each element containing unperturbed photosphere, dark spots, and facular areas. The fraction of the $k$-th  element covered by dark spots is indicated by its filling factor $f_{k}$,  the fractional  area of the faculae is $Qf_{k}$, and the fractional area of the unperturbed photosphere is $1-(Q+1)f_{k}$. 
The contribution to the stellar flux coming from the $k$-th surface element at time $t_{j}$, where $j=1,..., N$,  is an index numbering the $N$  points along the light curve, is given by:
\begin{eqnarray}
\Delta F_{kj} & = & I_{0}(\mu_{kj}) \left\{ 1-(Q+1)f_{k} + c_{\rm s} f_{k} +  \right. \nonumber \\
  & & \left.  Q f_{k} [1+c_{\rm f} (1 -\mu_{kj})] \right\} A_{k} \mu_{kj} {w}(\mu_{kj}),
\label{delta_flux}
\end{eqnarray}
where $I_{0}$ is the specific intensity in the continuum of the unperturbed photosphere at the isophotal wavelength of the observations, $c_{\rm s}$ and $c_{\rm f}$ are the spot and facular contrasts, respectively \citep[cf. ][]{Lanzaetal04}, $A_{k}$ is the area of the $k$-th surface element,
\begin{equation}
 {w} (\mu_{kj}) = \left\{ \begin{array}{ll} 
                      1  & \mbox{if $\mu_{kj} \geq 0$}  \\
                      0 & \mbox{if $\mu_{kj} < 0$ }
                              \end{array} \right. 
\end{equation}
is its visibility, and 
\begin{equation}
\mu_{kj} \equiv \cos \psi_{kj} = \sin i \sin \theta_{k} \cos [\ell_{k} + \Omega (t_{j}-t_{0})] + \cos i \cos \theta_{k},
\label{mu}
\end{equation}
is the cosine of the angle $\psi_{kj}$ between the normal to the surface element and the direction of the observer, with $i$ being the inclination of the stellar rotation axis to the line of sight, $\theta_{k}$ the colatitude and $\ell_{k}$ the longitude of the $k$-th surface element, $\Omega$ the angular velocity of rotation of the star ($\Omega \equiv 2 \pi / P_{\rm rot}$),  and $t_{0}$ the initial time. The specific intensity in the continuum varies according to a quadratic limb-darkening law, as adopted by \citet{Lanzaetal03} for the case of the Sun, viz. $I_{0} \propto a_{\rm p} + b_{\rm p} \mu + c_{\rm p} \mu^{2}$. The stellar flux computed at time $t_{j}$ is then: $F(t_{j}) = \sum_{k} \Delta F_{kj}$. To warrant a relative precision of the order of $10^{-5}$ in the computation of the flux $F$, each surface element is further subdivided into elements of $1^{\circ} \times 1^{\circ}$ and their contributions, calculated according to Eq.~(\ref{delta_flux}), are summed up at each given time to compute the contribution of the $18^{\circ} \times 18^{\circ}$ surface element to which they belong.  

We fit the light curve by changing the value of the spot filling factor $f$ over the surface of the star, while $Q$ is held constant. Even fixing the rotation period, the inclination, and the spot and facular contrasts \citep[see ][ for details]{Lanzaetal07}, the model has 200 free parameters and suffers from  non-uniqueness and instability. To find a unique and stable spot map, we apply ME regularization, as described in \citet{Lanzaetal07}, by minimizing a functional $Z$, which is a linear combination of the $\chi^{2}$ and  the entropy functional $S$; i.e.,
\begin{equation}
Z = \chi^{2} ({\vec f}) - \eta S ({\vec f}),
\end{equation}
where ${\vec f}$ is the vector of the filling factors of the surface elements, $\eta > 0$   a Lagrangian multiplier determining the trade-off between light curve fitting and regularization, and { the expression of $S$ is: 
\begin{equation}
S = -\sum_{k} w_{k} \left[ f_{k} \log \frac{f_{k}}{m} + (1-f_{k}) \log \frac{1-f_{\rm k}}{1-m} \right],
\end{equation}
where $w_{k}$ is the relative area of the $k$-th surface element (total surface area of the star $=1$) and $m$ the default spot covering factor that fixes the limiting values for $f_{k}$ as: $m < f_{k} < (1-m)$. In our modelling we adopt $m=10^{-6}$   \citep[cf. ][]{Lanzaetal98}. } The entropy functional $S$ is constructed in such a way that it attains its maximum value when the star is immaculate. Therefore, by increasing the Lagrangian multiplier $\eta$, we increase the weight of $S$ in the model and the area of the spots  is progressively reduced.
This gives rise to systematically negative residuals between the observations and the best-fit model when
$\eta > 0$. 

The optimal value of $\eta$ depends on the information content of the light curve, which in turn depends on the ratio of the amplitude of its rotational modulation to the average standard deviation of its  points. In the case of HD~189733, the  amplitude of the  rotational modulation is $\sim 0.015$, while the nominal standard deviation of the  points is $\sim 1.15 \times 10^{-4}$ in relative flux units (see Sect.~\ref{photometry}), giving a signal-to-noise ratio of $\sim 130$. However, since short-term instrumental fluctuations cannot be excluded in the case of  MOST photometry (cf. Sect.~\ref{photometry}), we prefer to consider the standard deviation of the residuals obtained with the unregularized best fit (i.e., with $\eta =0$; see Sect.~\ref{light_curve_model}). In such a way, we find that the relative accuracy of  MOST photometry is overestimated by at least a factor of $\sim 3$, i.e. the mean standard deviation of the photometric points is  $\sim 4 \times 10^{-4}$, leading to a signal-to-noise ratio of $\sim 40-50$. 

To fix the optimal value of the Lagrangian multiplier $\eta$, we compare the modulus of the mean of the residuals of the regularized best fit $|\mu_{\rm reg}|$ with the standard error of the residuals themselves, i.e., $\epsilon_{0} \equiv \sigma_{0}/\sqrt{N}$, where $\sigma_{0}$ is the standard deviation of the residuals of the unregularized best fit and $N$ is the number of points in each fitted subset of the light curve having a duration  $\Delta t_{\rm f}$ (see below). We iterate until $|\mu_{\rm reg}| \simeq \beta \epsilon_{0}$, where $\beta$ is a numerical parameter that will be fixed a posteriori according to the requisites of an acceptable quality for the fit and a regular evolution of the spot pattern from one spot map to the next \citep[cf. the  case of \object{CoRoT-4} in ][]{Lanzaetal09b}. 

In the case of the Sun, by assuming a fixed distribution of the filling factor, it is possible to obtain a good fit of the irradiance changes only for a limited time interval $\Delta t_{\rm f}$, not exceeding 14 days, which is the lifetime of the largest sunspot groups dominating the irradiance variation \citep{Lanzaetal03}. In the case of other active stars, the value of $\Delta t_{\rm f}$ must be determined from the observations themselves, looking for the maximum data extension that allows us a good fit with the applied model (see Sect.~\ref{model_param}). 

The optimal values of the spot and facular contrasts and of the facular-to-spotted area ratio $Q$ in stellar active regions are  unknown a priori. In our model the facular contrast $c_{\rm f}$ and the parameter $Q$ enter as the product $c_{\rm f} Q$, so we can fix $c_{\rm f}$ and vary $Q$, estimating its best value 
 by minimizing the $\chi^{2}$ of the model, as shown in Sect.~\ref{model_param}. Since the number of free parameters of the ME model is large,  we make use of the model of \citet{Lanzaetal03} { to fix the value of $Q$. It fits } the light curve by assuming only three active regions to model the rotational modulation of the flux plus a uniformly distributed background to account for the variations of  the mean light level. This procedure is the same as adopted  for \object{CoRoT-2} and \object{CoRoT-4} to fix the value of $Q$ 
\citep[cf. ][]{Lanzaetal09a,Lanzaetal09b}.  

We shall assume an inclination of the rotation axis of HD~189733 of $ i \simeq 85^{\circ}$ (see Sect.~\ref{model_param}). Since the information on spot latitudes that can be extracted from the rotational modulation of the flux for such a high inclination is negligible, the ME regularization virtually puts all the spots at the sub-observer latitude (i.e., $90^{\circ} -i \approx 5^{\circ}$) to minimize their area and maximize the entropy. Therefore, we are limited to mapping  only the distribution of the active regions vs. longitude, which can be done with a resolution of at least   $\sim 50^{\circ}$ \citep[cf. ][]{Lanzaetal07,Lanzaetal09a}. Our ignorance of the true facular contribution to the light modulation may lead  to systematic errors in the active region longitudes derived by our model, as discussed  by \citet{Lanzaetal07} in the case of the Sun.

\subsection{Modelling the line distortion induced by activity}
\label{line_prof_model}

Surface magnetic fields in late-type stars produce brightness and convection inhomogeneities that distort their spectral line profiles leading to apparent RV variations \citep[cf., e.g., ][]{SaarDonahue97,Saaretal98}. To compute the apparent RV variations induced by stellar active regions, we adopt a simple model for each line profile considering the residual profile $R(\lambda)$ at a wavelength $\lambda$ along the line, i.e.: $R(\lambda) \equiv [1 - I(\lambda)/I_{\rm c}]$, where $I(\lambda)$ is the specific intensity at wavelength $\lambda$ and $I_{\rm c}$ the intensity in the  continuum adjacent to the line. The local residual profile is assumed to be a Gaussian  with thermal and microturbulent width 
$\Delta \lambda_{\rm D}$ \citep[cf., e.g., ][ Ch.~1]{Gray88}, i.e.:
\begin{equation}
R(\lambda) \propto \exp \left[- \left( \frac{\lambda -\lambda_{0}}{\Delta \lambda_{\rm D}} \right)^{2} \right],
\label{local_prof}
\end{equation} 
where $\lambda_{0}$ is the local central wavelength. The different effective temperatures in dark spots or bright faculae modify the depth of the local line profile. To take into account this effect, we multiply the unperturbed profile given by Eq.~(\ref{local_prof}) by $(1-C_{\rm d}f_{k})$, where  $C_{\rm d}$ is a coefficient that  depends on the specific line considered and the physical properties of the stellar atmosphere, and $f_{k}$ is the spot filling factor in the specified surface element. This factor accounts for the temperature reduction in dark spots, but does not consider the increased effective temperature in faculae because the facular contrast is assumed to be solar-like (cf. Sect.~\ref{model_param}), i.e., corresponding to a modest temperature increase of $\approx 100$~K in the layers where the optical spectrum is produced \citep[cf., e.g.,  ][]{Unruhetal99}. 
Since the RV measurement is based on a cross-correlation function, $C_{\rm d}$ must be averaged over a great number of spectral lines having in general quite different temperature responses. For this reason, we  consider $C_{\rm d}$ as a free parameter in our model and adjust it to fit the observed RV variations induced by stellar activity  (see below for the effect of the variation of $C_{\rm d}$ on the RV perturbation induced by a starspot). 
\begin{figure}[t]
\centerline{
\includegraphics[width=6cm,height=8cm,angle=0]{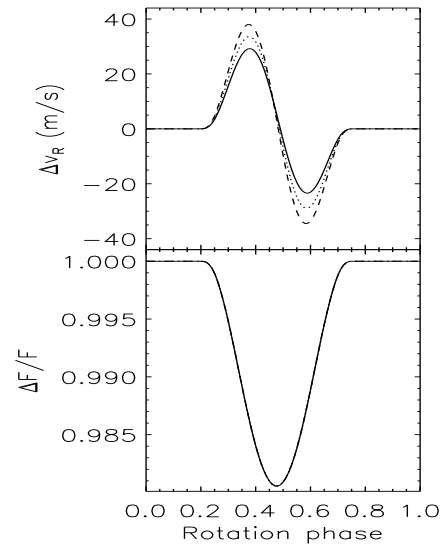}}
\caption{{\it Upper panel:} The RV variation produced by a single equatorial spot on HD~189733 vs. the rotation phase for different values of the parameter $C_{\rm d}$ according to the different linestyles: $C_{\rm d}=0.0$ (no temperature effect on the line profile): solid line; $C_{\rm d}=0.2$: dotted line; and $C_{\rm d}=0.4$: dashed line. {\it Lower panel:} the corresponding modulation of the optical flux vs. the rotation phase. Note that the three plots overlap because the modulation is independent of $C_{\rm d}$ (see the text). 
}
\label{cd_effect}
\end{figure}
In addition to the variation of the local line profile, the local specific intensity along a spectral line $I(\lambda)$ depends  on the variation of the continuum intensity $I_{\rm c}$ owing to limb-darkening and the effects of dark spots and bright faculae, as specified by  Eq.~(\ref{delta_flux}).  

To include the effects of surface magnetic fields on convective motions, we consider 
the decrease of macroturbulence velocity  and the reduction of convective blueshifts in active regions.
Specifically, we assume a local radial-tangential macroturbulence, as introduced by \citet{Gray88}, with a distribution function $\Theta$ for the radial velocity $v$ of the form: 
\begin{equation}
\Theta (v, \mu) = \frac{1}{2} \frac{\exp \left[- \left( \frac{v}{\zeta_{\rm RT} \sqrt{1 -\mu^{2}}} \right)^{2} \right]}{\sqrt{\pi} \zeta_{\rm RT} \sqrt{1 - \mu^{2}}} + \frac{1}{2} \frac{\exp \left[- \left( \frac{v}{\zeta_{\rm RT} \mu} \right)^{2} \right]}{\sqrt{\pi} \zeta_{\rm RT} \mu}, 
\label{macroturb}
\end{equation}
where $\zeta_{\rm RT}$ is the macroturbulence and $\mu$ is given by Eq.~(\ref{mu}). The  profile emerging from each surface element is obtained by convolving the local profile in Eq.~(\ref{local_prof}) with the  Doppler shift distribution as generated by the macroturbulence function in Eq.~(\ref{macroturb}). 
The parameter $\zeta_{\rm RT}$ is assumed to be  reduced in spotted and facular areas, according to their total filling factor, i.e., $\zeta_{\rm RT} = \zeta_{0} [1- (Q+1)f]$, where $\zeta_{0}$ is the unperturbed macroturbulence.  This is a convenient parameterization of the quenching effect of surface magnetic fields  on convective turbulence, at least in the framework of our simplified model, and is supported by the observations of the reduction of turbulent velocity fields in the plage regions of the Sun \citep[cf., e.g., ][]{Titleetal92}.  

Convective blueshifts arise because in stellar photospheres most of the flux comes from the extended updrafts at the centre of  convective granules. At the centre of the stellar disc, 
the vertical component of the convective velocity  produces a maximum blueshift, while
the effect vanishes at the limb where the projected  velocity is zero. 
The cores of the deepest spectral lines form in the upper layers of the photosphere where the vertical convective velocity is low, while the cores of shallow lines form in deeper layers with higher  vertical velocity. Therefore, the cores of shallow lines are blueshifted with respect to the cores of the deepest lines. \citet{Gray09} shows this effect by plotting the endpoints of line bisectors of shallow and deep lines on the same velocity scale. He shows that the amplitude of the relative blueshifts scales with the spectral type and the luminosity class of the star. For the G8V star \object{$\tau$ Ceti}, which has a spectral type not too much different from the K1V of HD~189733, the convective blueshifts are approximately  similar to those of the Sun, so we adopt solar values in our simulations. 
In active regions, vertical convective motions are quenched, so we observe an apparent redshift of the spectral lines in spotted and facular areas in comparison to the unperturbed photosphere. \citet{Meunieretal10b}  quantify this effect in the case of the Sun, and we adopt their results, considering an apparent redshift $\Delta V_{\rm f}=200$ m~s$^{-1}$ in faculae and $\Delta V_{\rm s} = 300$ m~s$^{-1}$ in cool spots. 

In principle,  the integrated effect of convective redshifts can be measured  in a star by comparing RV measurements obtained with two different line masks, one including the shallow lines and the other the  deep lines \citep[cf. ][]{Meunieretal10b}. For HD~189733, which lacks  such  measurements, we apply the results of  \citet{Gray09} and adopt solar-like values as the best approximation. 

Considering solar convection as a template, intense downdrafts are localized in the dark lanes at the boundaries of the upwelling granules, but they contribute a significantly smaller flux because of their lower brightness and smaller area. While a consideration of those downdrafts is needed to simulate the shapes of line bisectors, it is beyond the scope of our simplified approach that assumes that the whole profile of our template line forms at the same depth within the photosphere.
Therefore,  we restrict our model to the mean apparent RV variations by  neglecting the associated variations of the bisector shape and do not include the effect of convective downdrafts,  as well as those of other surface flows typical of solar active regions, such as the Evershed effect in sunspots \citep[cf., e.g., ][]{Meunieretal10b}.   

The local central wavelength $\lambda_{0}$ of the $k$-th surface element at time $t_{j}$ is given by  
$\lambda_{0kj} = \lambda_{\rm R} (1+ v_{kj}/c)$, where $v_{kj}$ is its radial velocity,  
\begin{equation}
v_{kj} = - (v \sin i) \sin \theta_{k} \sin [\ell_{k} + \Omega  (t_{j} - t_{0})] + \Delta V_{\rm cs},
\label{rv_perturb}
\end{equation}
with $c$ the speed of light, 
 $v \sin i$  the stellar projected rotational velocity,  $\lambda_{\rm R}$ the rest wavelength of the spectral line, and $\Delta V_{\rm cs}$ the apparent convective redshift arising from the reduction of convective blueshifts in spots and faculae, which is parameterized as 
\begin{equation}
\Delta V_{\rm cs} = ( \Delta V_{\rm s}  + Q  \Delta V_{\rm f}) f_{k} \mu_{kj},  
\label{cshifts}
\end{equation}
where $f_{k}$ is the spot filling factor of the element, $Q$ the facular-to-spotted area ratio, and $\mu_{kj}$ is given by Eq.~(\ref{mu}). 

We integrate the line specific intensity at a given wavelength over the disc of the star using a subdivision into  $1^{\circ} \times 1^{\circ}$ surface elements to obtain the flux along the line profile $F(\lambda, t_{j})$ at a given time $t_{j}$. To find the apparent stellar RV, we can fit a Gaussian to the  line profile $F(\lambda, t_{j})$, or we can determine the centroid of the profile as 
\begin{equation}
 \lambda_{\rm ce} (t_{j}) = \frac{\int \lambda {\cal R}(\lambda, t_{j})\, d \lambda}{\int {\cal R}(\lambda, t_{j})\, d \lambda}, 
\label{centroid}
\end{equation}
where ${\cal R} \equiv [1-F(\lambda, t_{j})/F_{\rm c} (t_{j})]$ is the residual profile of the spectral line computed from the ratio of the flux in the line $F(\lambda, t_{j})$ to the flux $F_{\rm c} (t_{j}) $ in the adjacent continuum at any given time $t_{j}$. For HD~189733, the two methods lead to almost equal values of the RV perturbation with  differences never exceeding $\sim 3$ percent. 

A single line profile computed with the above model can be regarded as a cross-correlation function ({hereafter} CCF) obtained by cross-correlating the whole stellar spectrum with a line mask consisting of Dirac delta functions  giving the rest wavelength and depth of each individual line \citep[e.g., ][]{Quelozetal01}. Therefore,  to derive the RV from a single synthetic line profile is equivalent to measuring the RV from a  CCF, either by fitting it with a Gaussian or by computing its centroid wavelength.  A better approach would be to  simulate the whole stellar optical spectrum or an extended section of it  to account for the wavelength  dependence of the spot and facular contrasts, as well as of the convective inhomogeneities \citep[cf., e.g.,  ][]{Desortetal07,Lagrangeetal10,Meunieretal10b}. Again, in view of our limited knowledge of the distribution of active regions on the stellar surface, we believe that our simplified approach is adequate for estimating the RV perturbations induced by magnetic fields. 

An illustration of the effects of dark spots, bright faculae, and macroturbulence perturbations on the stellar RV measurements is presented in \citet{Lanzaetal10} together with a comparison of the results of the present model with those of previous approaches. { Here we only show in Fig.~\ref{cd_effect} (upper panel) the effect of the variation of the parameter $C_{\rm d}$ on the RV curve of a single spot that was not considered by \citet{Lanzaetal10}. The adopted spot is located on the equator of the star and has a square shape with a side of $18^{\circ}$, a covering factor $f=0.99$,  and a contrast $c_{\rm s}=0.482$ without any facular contribution. The stellar and line profile parameters are assumed to be the same as adopted for the modelling of the RV variations of HD~189733 (see Sect.~\ref{model_param}). The increase of the amplitude of the RV variation with $C_{\rm d}$ is a consequence of the higher bump associated with the spot on the line profile when the depth of the local profile is reduced by the factor $(1-C_{\rm d}f)$. On the other hand, the modulation of the optical flux associated with the transit of the spot across the stellar disc is independent of the value of $C_{\rm d}$ because it depends on the flux in the continuum outside the spectral line (cf. Fig.~\ref{cd_effect}, bottom panel).}

The amplitude of the RV variation produced by an active region depends on several parameters that are poorly known, i.e., the latitude of the active region, the spot and facular contrasts, and the macroturbulence parameter which is difficult to  disentangle from rotational broadening in a slowly rotating star such as HD~189733 \citep[][]{Legeretal09}. Moreover, the spot and facular contrasts depend on the wavelength \citep{Lanzaetal04}, leading to a difference of $\approx 10$ percent in the RV variations as derived from different orders of an echelle spectrum \citep[cf., e.g., ][]{Desortetal07}. The simultaneous presence of several active regions gives rise to a complex line profile distortion in the case of a slowly rotating star because the perturbation of each active region is not separated from the other by the rotational broadening. This implies an additional $10-15$ percent uncertainty in the determination of the central wavelength of the profile by the Gaussian fit or the centroid method, even for an infinite signal-to-noise ratio and perfectly regular sampling, as in the case of simulated line profiles. In consideration of all these limitations, the absolute values of the RV variations computed with our model are uncertain by $10-20$ percent in the case of complex distributions of active regions, such as those derived by our spot modelling technique as applied to MOST and CoRoT light curves.      

\subsection{RV variations from spot modelling}
\label{rv_from_model}

We can use the distribution of active regions  as derived from our  spot modelling to synthesize the corresponding RV variations according to the approach outlined in Sect.~\ref{line_prof_model}. Since our spot model assumes that active regions are stable for the time interval of each fitted time series $\Delta t_{\rm f}$, the distribution of surface inhomogeneities  can be used to synthesize the RV variations having a timescale of $\Delta t_{\rm f}$ or longer. To improve the time resolution, we shall consider models obtained from time intervals whose initial time is shifted by $\Delta t_{\rm f}/2$ from the previous interval, and average the RV variations computed at the same time  to account for the spot evolution in a simple way.
Our ME modelling minimizes the area of the starspots and puts them close to the equator. Since the spot latitudes cannot  be determined from photometry  (cf. Sect.~\ref{spotmodel}), we assume that all the spots are located close to  the equator, which maximizes the computed RV variations (cf. Eq. \ref{rv_perturb}). { Note that the frequency content of the simulated RV time series is not significantly modified by this assumption because the duration of the transit of a given spot across the stellar disc is pratically independent of its latitude given that the inclination of the stellar rotation axis is close to $90^{\circ}$ and our reference frame is rigidly rotating with the mean stellar rotation period. Moreover, the occurrence times of the maximum and minimum of the RV variations associated with a given spot are mainly determined by its longitude and its projected area that our modelling approach consistently retrieves from the modulation of the optical flux. Therefore, we do not expect any significant modification of the power spectrum of the synthetic RV variations associated with the assumption of equatorial spots. }

The active regions with  lifetimes shorter than  $\Delta t_{\rm f}/2$ produce a photometric modulation that  appears in the residuals of the best fit to the lightcurve. As we shall see in Sect.~\ref{light_curve_model}, most of the short-term variability occurs on timescales of $1-2$ days, i.e., significantly shorter than the rotation period of HD~189733, so we can neglect, as a first approximation, the variation in the disc position of  active regions due to stellar rotation and estimate their area as if they were located at the centre of the disc. 
Of course, the actual RV perturbation depends on the position of the active regions on the disc, but this information is lacking. Therefore, the approach we describe below cannot be used to synthesize the actual RV variation, but only to estimate its maximum amplitude. 

First, we express the photometric residuals as the relative deviation $\Delta F/F_{0}$ between the observed flux and its best fit measured in units of the reference unspotted flux $F_{0}$. Then we subtract  its mean value $\mu_{\rm res} \equiv \langle \Delta F /F_{0} \rangle$ from $\Delta F/F_{0}$ because the mean value corresponds to a uniformly distributed population of active regions that does not produce any RV variation. 
At each given time $t$, we adopt  $| \Delta F/F_{0} - \mu_{\rm res}| c_{\rm s}^{-1} $ as a measure of the filling factor of the  active regions producing the short-term RV variations. In doing so, we neglect limb-darkening effects and assume that those active regions  consist of  dark spots with a contrast $c_{\rm s}$. Finally, we compute the amplitude of the RV perturbation due to such brightness inhomogeneities by means of Eq.~(1) of \citet{Desortetal07} obtaining  
\begin{equation}
\Delta V_{\rm Ri} = 800 \left| \frac{\Delta F}{F} - \mu_{\rm res} \right| c_{\rm s}^{-1} (v \sin i)^{1.1},
\label{rv_short_term}
\end{equation}
where the RV perturbation is measured in m~s$^{-1}$ and the $v \sin i$ of the star in km~s$^{-1}$. Eq.~(\ref{rv_short_term}) gives an upper limit for the amplitude of the RV perturbation because it assumes that an active region spans the diameter of the stellar disc.  
 On the other hand, the effects of  convective redshifts in short-lived active regions is estimated as 
\begin{equation}
\Delta V_{\rm Rc} = (\Delta V_{\rm s } + Q \Delta V_{\rm f}) \left| \frac{\Delta F}{F} - \mu_{\rm res} \right| c_{\rm s}^{-1}.  
\label{rv_short_csh}
\end{equation}    
By adding both short-term perturbations, we compute the amplitude of the total variation induced by the active regions evolving on time scales shorter than the time resolution of our model, and add it to the standard deviation of the RV variations computed with the slowly evolving active regions to compute the total uncertainty of the synthesized RV values.

\section{Model parameters}
\label{model_param}

The basic stellar parameters are taken from \citet{Bouchyetal05} and \citet{Triaudetal09} and are listed in 
Table~\ref{model_param_table}.  The limb-darkening parameters have been derived from \citet{Kurucz00} model atmospheres for $T_{\rm eff} = 5050$~K, $\log g = 4.53$~(cm s$^{-2}$) and solar abundances, by adopting the MOST transmission profile in \citet{Walkeretal03}. We compared our limb-darkening profile with that derived by  \citet{Miller-Riccietal08}, who adopted a different limb-darkening law and used MOST data acquired in 2006,  and found a maximum relative difference of 3 percent. 

The rotation period adopted for our spot modelling  has been derived from  the optical light curve  ($P_{\rm rot} = 11.953 \pm 0.009$ days) by \citet{HenryWinn08}.  
\begin{table}
\noindent 
\caption{Parameters adopted for the light curve and radial velocity modelling of HD~189733.}
\centering
\begin{tabular}{lrr}
\hline
 & & \\
Parameter & Value & Ref.$^{a}$\\
 & & \\ 
\hline
 & &  \\
Star Mass ($M_{\odot}$) & 0.823 & T09  \\
Star Radius ($R_{\odot}$) & 0.766 & T09  \\
$T_{\rm eff}$ (K) & 5050 &  B05 \\
$\log g$ (cm s$^{-2}$) & 4.53 & B05 \\ 
$a_{\rm p}$ & 0.213 & L11 \\
$b_{\rm p}$ & 0.933 & L11 \\
$c_{\rm p}$ & -0.152 & L11 \\ 
$P_{\rm rot}$ (days) & 11.953 & HW08 \\
$\epsilon$ & $8.54 \times 10^{-6}$ & L11 \\ 
Inclination (deg) & 85.3 & B05, T09  \\
$c_{\rm s}$  & 0.482 & L11 \\
$c_{\rm f}$  & 0.115 & L04 \\ 
$Q$ & 0.0  & L11 \\ 
$\Delta t_{\rm f}$ (days) & 7.625 & L11 \\ 
$v \sin i$ (km s$^{-1}$) & 3.316 & T09 \\
$\Delta \lambda_{\rm D} $ (km s$^{-1}$) &  2.33 & L11 \\
$\zeta_{0}$ (km s$^{-1}$) & 1.2 & L11 \\  
& &   \\
\hline
\label{model_param_table}
\end{tabular}
~\\
$^{a}$ References: B05: \citet{Bouchyetal05}; HW08: \citet{HenryWinn08}; L04: \citet{Lanzaetal04}; L11: present study; T09: \citet{Triaudetal09}. 
\end{table}
The polar flattening of the star due to the centrifugal potential is computed in the Roche approximation with a rotation period of 11.95 days. The relative difference between the equatorial and the polar radii is $\epsilon = 2.58 \times 10^{-5}$ which induces a completely negligible relative flux variation of $\approx 10^{-6}$  for a spot coverage of $\sim 2$ percent, as a consequence of the gravity darkening of the equatorial regions of the star. 

The inclination of the stellar rotation axis is constrained by the observations of the Rossiter-McLaughlin effect by \citet{Triaudetal09} who found that the angle $\beta_{\rm RM}$ between the projected directions on the plane of the sky of the orbital angular momentum and the stellar spin  is $\beta_{\rm RM} = 0 \fdg 85 \pm 0 \fdg 32$. Although the possibility that the stellar spin be not aligned with the orbital angular momentum cannot be excluded with certainty because only the sky-projected angle between the two vectors has been measured, we regard this occurrence as highly improbable and assume that the rotation axis is inclined with respect to the line of sight of the same angle as the orbit normal.  

The maximum time interval that our model can accurately fit with a fixed distribution of active regions $\Delta t_{\rm f}$ has been determined by dividing the total interval, $T= 30.4994$ days, into $N_{\rm f}$ equal segments, i.e., $\Delta t_{\rm f} = T/N_{\rm f}$, and looking for the minimum value of $N_{\rm f}$ that allows us a good fit of the light curve, as measured by the $\chi^{2}$ statistics. We found that for $N_{\rm f} < 4$ the quality of the best fit degrades significantly with respect to higher values, owing to a  substantial evolution of the pattern of surface brightness inhomogeneities. We adopt $N_{\rm f} = 4$, and therefore  $\Delta t_{\rm f} = 7.625$ days is the maximum time interval to be fitted with a fixed distribution of surface active regions in order to  estimate the best value of the parameter $Q$ (see below). 

To compute the spot contrast, we adopt the same mean temperature difference as derived for sunspot groups from their bolometric contrast, i.e. 560~K \citep{Chapmanetal94}. In other words, we assume a spot effective temperature of $ 4490$~K, yielding a contrast $c_{\rm s} = 0.482$ in the MOST passband \citep[cf. ][]{Lanzaetal07}.
A different spot contrast changes the absolute spot coverages, but does not significantly affect their longitudes and their time evolution, as discussed in detail by \citet{Lanzaetal09a}. Therefore, adopting a spot temperature deficit of, e.g., $\sim 1000$~K, as discussed in the analysis of HD~189733 HST data by \citet{Pontetal07}, will decrease the filling factor in proportion to the lower contrast, but does not change our results in any significant way. 

\begin{figure}[t]
\centerline{
\includegraphics[width=6cm,height=8cm,angle=90]{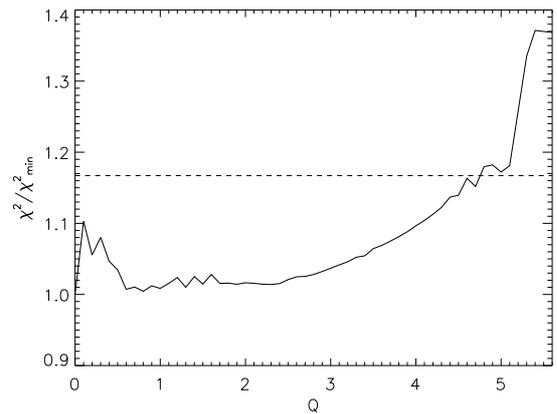}} 
\caption{
The ratio of the $\chi^{2}$ of the composite best fit of the entire time series of HD~189733  
to its minimum value vs. the parameter $Q$, i.e., the ratio of the area of the faculae to that of the cool spots in active regions. The horizontal dashed line indicates the 95 percent confidence level for $\chi^{2}/\chi_{\rm min}^{2}$, determining the interval of acceptable $Q$ values.
}
\label{qratio}
\end{figure}

In the models computed with a facular contribution, the facular contrast is assumed to be solar-like with $c_{\rm f} = 0.115$ \citep{Lanzaetal04}.    
The best value of the area ratio $Q$ between the faculae and the spots in each active region has been estimated by means of the three-spot model by \citet[][ cf. Sect.~\ref{spotmodel}]{Lanzaetal03}. In Fig.~\ref{qratio}, we plot the ratio $\chi^{2}/ \chi^{2}_{\rm min}$ of the total $\chi^{2}$ of the composite best fit of the entire time series to its minimum value $\chi^{2}_{\rm min}$, versus $Q$, and indicate the 95 percent confidence level as derived from the F-statistics \citep[e.g., ][]{Lamptonetal76}. The choice of $\Delta t_{\rm f} = 7.625$~days allows us to fit the rotational modulation of the active regions for the longest time interval during which they remain stable, modelling both the flux increase due to the facular component when an active region is close to the limb as well as the flux decrease due to the dark spots when the same region transits across the central meridian of the disc. In such a way, a measure of the relative facular and spot contributions can be obtained leading to an  estimate of $Q$. 
Fig.~\ref{qratio} shows that the best value of $Q$ is $Q=0$, with an acceptable range extending from~$\sim 0$~to~$\sim 5$. { Therefore, we adopt $Q=0$ for all our modelling in Sect.~\ref{results} unless otherwise indicated. We shall further comment on the value of $Q$ in Sect.~\ref{conclusions}. } 

To compute the RV variations induced by surface inhomogeneities, we assume a line rest wavelength of $600$~nm and a local thermal plus microturbulence broadening $ \Delta \lambda_{\rm D} = 2.3$~km~s$^{-1}$.  The  $v \sin i = 3.316 \pm 0.07$~km~s$^{-1}$ is derived from the observations of the Rossiter-McLaughlin effect \citep{Triaudetal09}. Although such a measurement can be affected by a systematic error, we prefer to use this value, that is larger than the $2.97 \pm 0.22$~km~s$^{-1}$ adopted by \citet{Winnetal06}, to maximize the synthesized RV variations. 
{ Adopting $v \sin i = 2.97$~km~s$^{-1}$ would imply an amplitude of the RV variations lower by $\approx 10$ percent for the same value of the parameter $C_{\rm d}$. On the other hand, we could still reproduce the observed variations by changing the average value of $C_{\rm d}$ (see  Sect.~\ref{results_rv}), so that the choice of the
$v \sin i$ value is not very critical for our modelling.} 
 The radial-tangential macroturbulence velocity is assumed $\zeta_{0} = 1.2$~km~s$^{-1}$ after \citet{Gray88} because it is difficult to disentangle  from the rotational broadening owing to the slow rotation of the star. 

\section{Results}
\label{results}

\subsection{Light curve models}
\label{light_curve_model}

We applied the model of Sect.~\ref{spotmodel} to the out-of-transit  light curve of HD~189733, considering time  intervals  $\Delta t_{\rm f} = 7.625$ days. The initial epoch of each fitted interval is shifted by $\Delta t_{\rm f}/2= 3.81$ days, to sample the evolution of the starspots in a more continuous way.  
The best fit without regularization ($\eta = 0$) has a mean  $\mu_{\rm res} = 2.15 \times 10^{-6}$ and a standard deviation of the residuals $\sigma_{0} = 3.99 \times 10^{-4}$ in relative flux units. The Lagrangian multiplier $\eta$ is iteratively adjusted until the mean of the residuals $\mu_{\rm res} = -7.82 \times 10^{-5} \simeq 2 \sigma_{0}/\sqrt{N}$, where $N  =  103$ is the mean number of  points in each fitted light curve interval  $\Delta t_{\rm f}$; the standard deviation of the residuals of the regularized best fit is $\sigma = 4.86 \times 10^{-4}$. In other words, we fixed $\beta \simeq 2$ in the regularization procedure which gives the best compromise between the accuracy of the fit and the smoothness of the spot distribution and evolution. 

\begin{figure*}[]
\centerline{
\includegraphics[width=12cm,height=18cm,angle=90]{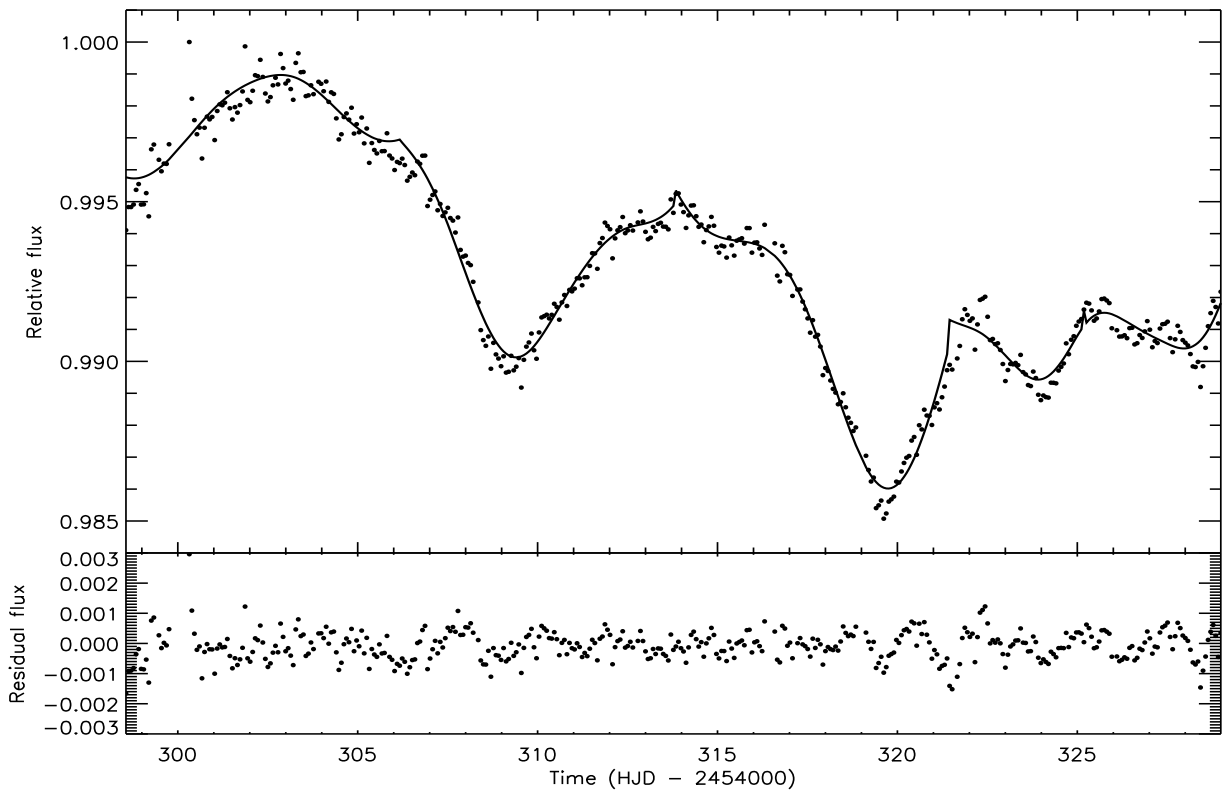}} 
\caption{{\it Upper panel:} The ME-regularized composite best fit to the out-of-transit light curve of HD~189733 obtained for $Q=0.0$. The flux is in relative units, i.e., measured with respect to the maximum observed flux along the light curve.  {\it Lower panel:} The residuals from the composite best fit versus time.
}
\label{lc_bestfit}
\end{figure*}

The composite best fit to the entire light curve is shown in the upper panel of Fig.~\ref{lc_bestfit} while the residuals are plotted in the lower panel. The computed flux values at the same time from successive spot models 
are averaged.  { In spite of that, at the matching points between successive best fits we observe some discontinuities in the first derivative of the flux variation vs. time. This may in principle affect the rate of change of the RV perturbation induced by the spots vs. the rotation phase. However, we do not worry about correcting for this effect because the discontinuities always occur at epochs when no RV observation has been obtained. Therefore, the synthesizes RV values to be compared with the SOPHIE RV time series are not appreciably affected by those discontinuities. 
}

The residuals show oscillations with a typical timescale of $\sim 1-2$ days that can be related to the rise and decay of  active regions that cover $\approx 0.2-0.3$ percent of the stellar disc, i.e., comparable with the largest sunspot groups. { The projected area of those active regions is estimated from the amplitude of the flux residuals and the adopted spot contrast, while their lifetimes are estimated by the duration of  the residual fluctuations. 
These active regions cannot be modelled by our approach because they do not produce a significant rotational flux modulation during the $\sim 12$ days of the stellar rotation period as they move across the disc by only $\approx 60^{\circ}$ in longitude. Their estimated size and lifetime are not significantly degenerate with each other because our model has 200 degrees of freedom to fit any minute  flux variation with a timescale of several days by adjusting the covering factors $f_{k}$ of its  $18^{\circ} \times 18^{\circ}$ surface elements. In other words, by subtracting  our model we filter out all the variations with time scales longer than $4-5$ days leaving only the effects of the active regions evolving on shorter timescales.}  

By decreasing the degree of regularization, i.e., the value of $\eta$, we can marginally improve the best fit, but at the cost of introducing several small active regions that wax and wane from one $\Delta t_{\rm f}$ time interval to the next and are badly constrained by the rotational modulation. Nevertheless, the oscillations of the residuals do not disappear completely even for $\eta = 0$, indicating that HD~189733 has a population of short-lived active regions with typical lifetimes of $1-2$ days. Spots can be mapped along the strip occulted by the planet, as shown by \citet{Pontetal07}, but this is beyond the scope of the present investigation. 

\subsection{Longitude distribution of active regions and stellar differential rotation}
\label{spot_model_res}

\begin{figure}
\centerline{
\includegraphics[width=8cm]{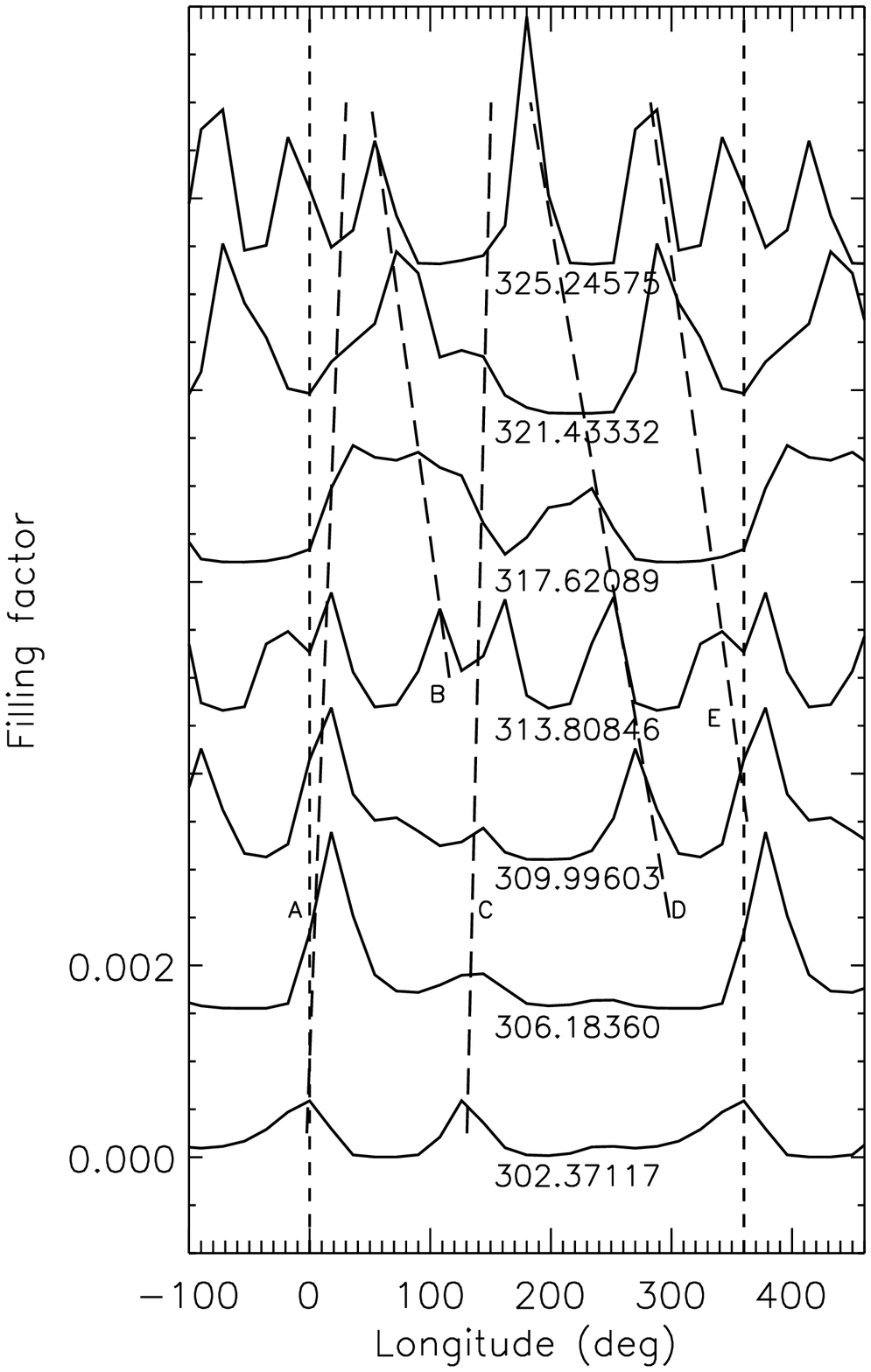}} 
\caption{The distributions of the spotted area vs. longitude at the labelled times (${\rm HJD}-2454000.0$) for $Q=0.0$. The plots have been vertically shifted to show the migration of individual spots (relative maxima of the distributions) versus time. The vertical short-dashed lines mark longitudes $0^{\circ}$ and $360^{\circ}$, beyond which the distributions have been repeated to help  following the  spot migration. The long-dashed lines labelled with capital letters from A to E trace the migration of the most conspicuous spots detected in the plots, respectively (see the text). 
}
\label{long_distr}
\end{figure}

The distributions of the spotted area vs. longitude are plotted in Fig.~\ref{long_distr}
for the seven mean epochs of our individual subsets adopting a rotation period of 11.953 days. The longitude zero corresponds to the point intercepted on the stellar photosphere by the line of sight to the centre of the star  at HJD~2454298.55874, i.e., the sub-observer point at the initial epoch. The longitude increases in the same direction as the stellar rotation. This is consistent with the reference frames adopted in our previous  studies \citep[][]{Lanzaetal09a,Lanzaetal09b}, but does not allow a direct comparison of the mapped active regions with the dips in the light curve. 

Our maps show that the distribution of the spotted area versus longitude evolves rapidly, which makes difficult to trace the evolution and migration of the active regions in an unambiguous way. Nevertheless, five main active regions can be identified in Fig.~\ref{long_distr} and their migration is traced with dashed lines labelled with letters from A to E. Their rotation rates  with respect to the adopted reference frame have  been estimated by a linear best fit, i.e., by assuming a constant migration rate, and 
range from $-5.9 \pm 1.5$ deg/day for that labelled with D to $1.1 \pm 1.5$ deg/day for that labelled with A. 
Note that  active longitudes D and E are not detectable in some of the ME distributions plotted in  Fig.~\ref{long_distr}, but we decided to trace their evolution in a continuous way attributing the change of their visibility to a combination of their fast intrinsic evolution and the differential rotation.  

The relative amplitude of the surface differential rotation, as estimated from the difference between the maximum and the minimum migration rates reported above, is $ \Delta \Omega /\Omega = 0.23 \pm 0.10$. Unfortunately, no information is available on the spot latitudes, thus our $\Delta \Omega / \Omega$ is only a lower limit. Since HD~189733 is more active than the Sun, in principle its active regions may cover a latitude range larger than in the Sun where sunspot groups are confined to $\pm 40^{\circ}$ from the equator \citep[see, e.g., ][]{Strassmeier09}. \citet{Faresetal10} using Zeeman Doppler Imaging and adopting a solar-like differential rotation profile, estimated equatorial and polar rotation periods of $11.94 \pm 0.16$ and $16.53 \pm 2.43$ days, giving a relative amplitude $\Delta \Omega / \Omega = 0.38 \pm 0.18$. Adopting their differential rotation profile with the above uncertainties, we estimate a latitude ranging from $40^{\circ}$ to $80^{\circ}$ for the highest latitude spots, i.e., those in the active longitudes B and D. On the other hand, spots in the active longitudes A and C are close to the equator and those in E are at intermediate latitudes. Since the  RV variations synthesized with all the spots located close to the equator reproduce the observations well (cf. Sect.~\ref{results_rv}), the lower value for the latitudes of the spots in B, D, and E, i.e., $\approx 30^{\circ}-40^{\circ}$, is favoured. Note that assuming, in the next sub-section, those spots close to the  equator leads to an overestimate of their effects in the induced RV variations of $\sim 20-25$ percent which is within our typical errorbars of $\pm 5$~m~s$^{-1}$. 

The spot longitudes in our model show in general a dependence on the facular-to-spotted area ratio $Q$ \citep[cf., e.g., ][]{Lanzaetal07}, therefore we computed regularized models with the maximum allowed facular contribution corresponding to $Q=5$. The relative amplitude of the differential rotation is not significantly modified with respect to the case with $Q=0$, i.e., it is still compatible with the result of \citet{Faresetal10}. 

\begin{figure}[b]
\centerline{
\includegraphics[width=6.5cm,angle=90]{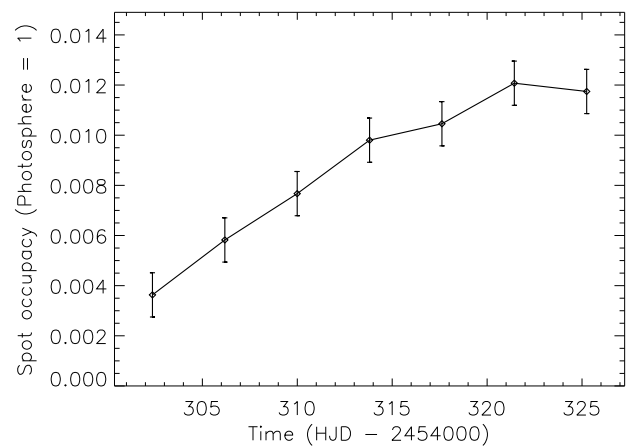}} 
\caption{The total spotted area as derived from the regularized ME models vs. time for $Q=0.0$. 
}
\label{total_area}
\end{figure}

The active longitudes traced in Fig.~\ref{long_distr} show significant changes on a timescale as short as $3-4$ days although their overall evolution may exceed the duration of the observations, i.e., $ \sim 30$ days.
The variation of the total spotted area vs. time is plotted in Fig.~\ref{total_area} and shows a monotonous increase along the MOST observing run corresponding to the overall decrease of the mean flux during the same interval. Note that the  area per longitude bin and  the total area depend on the spot and facular contrasts adopted for the modelling. Specifically, darker spots lead to a smaller total area while the effect of the facular contrast is more subtle and influences somewhat the longitudinal distribution of the active regions \citep[see, ][ for more details]{Lanzaetal07,Lanzaetal09a}. 

Finally, we note that the time resolution of our spot models is not adequate to look for a possible modulation of the stellar activity with the orbital period of its  close-in massive planet, as suggested by \citet{Shkolniketal08}. Given the short orbital period of the planet, a different approach should be used to search for signatures of a possible star-planet interaction, as in the case of CoRoT-2 \citep[cf., e.g., ][]{Paganoetal09}. 

\subsection{Activity-induced RV variations}
\label{results_rv}

To simulate the apparent RV changes induced by the distribution of active regions derived from our  light curve modelling, we considered a spectral line with a rest wavelength of $600$ nm that is close to the isophotal wavelength of MOST observations for which our contrast coefficients are given. The spectral  resolution of the line profile is $\lambda/\Delta \lambda = 75,000$, i.e., comparable to that of SOPHIE. 

We fit the observed RV residuals by adjusting an offset value and the parameter $\langle C_{\rm d} \rangle $ that measures the average reduction of the line depth inside starspots. We exclude from the fit the four measurements between HJD~2454308.0 and 2454310.0 because those residuals show a very steep variation with a decrease of the RV of  $\sim 25 $~m~s$^{-1}$ in only two days, that is impossible to reproduce with our model. 
This fast change and other similar rapid variations in the observed RV time series exceed the accuracy of the SOPHIE velocimetry (cf. Sect.~\ref{rv_measurements}). They could be due to photospheric velocity fields of several km~s$^{-1}$ localized in active regions and displaying short-term variations, such as those observed in the Sun { during flares} or the emergence of new magnetic flux, or in syphon flows \citep[e.g., ][]{Ruedietal92}, or in the  time-varying features of the Evershed flows \citep[e.g., ][]{CabreraSolanaetal07}. 
We looked at the variability of the residuals in the other published RV datasets  \citep{Bouchyetal05,Winnetal06}. They also show  RV changes on short time scales with night-to-night variations of several tens of m~s$^{-1}$, in agreement with the present SOPHIE observations  \citep[see, e.g., the bottom panel of Fig.~1 in ][]{Bouchyetal05}. 
{ The chromospheric activity, as monitored by the core emission of the Ca~II H\&K lines, shows remarkable variations on time scales as short as one hour, suggesting frequent flaring events that can contribute to the short-term RV variations \citep{Moutouetal07,Faresetal10}.}

\begin{figure*}[]
\centerline{
\includegraphics[width=12cm,height=18cm,angle=90]{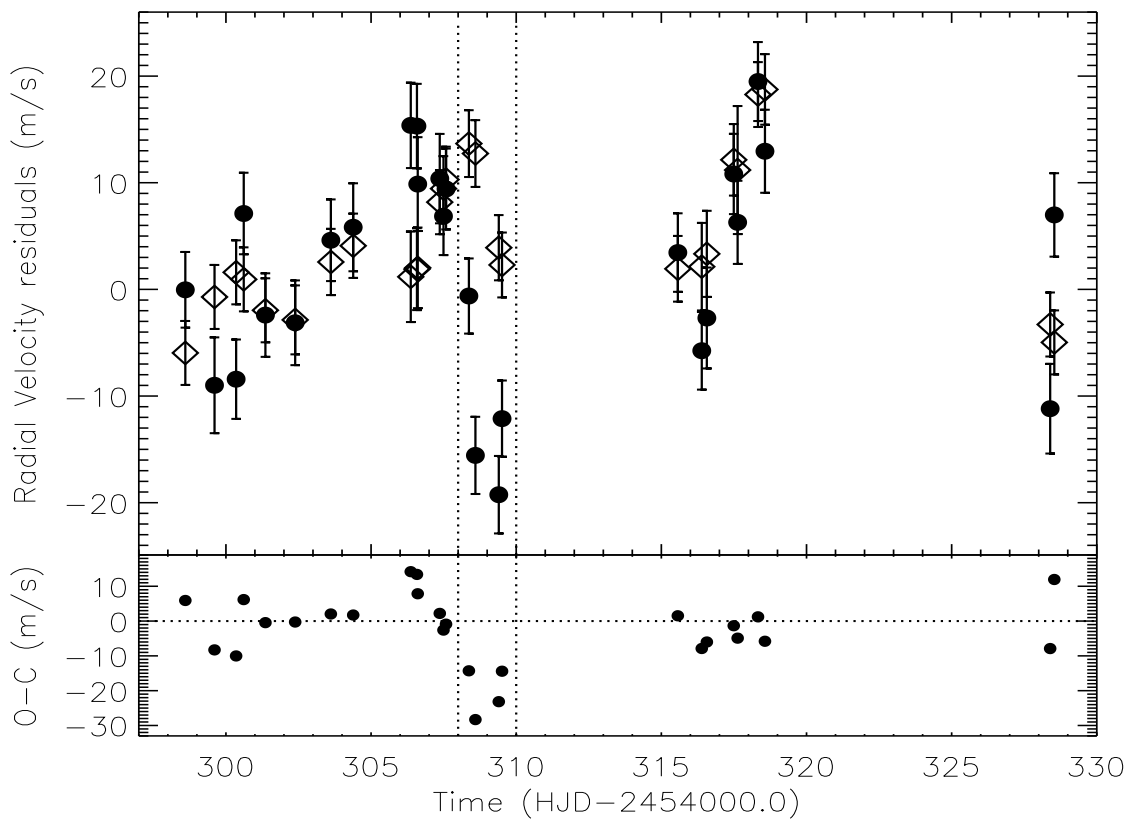}} 
\caption{{\it Upper panel:} The observed RV variations due to stellar activity (filled dots) and the  corresponding synthesized variations (open diamonds) versus time for the spot models with $Q=0$. The vertical dotted lines mark the time interval with the steepest variations that cannot be fitted by our model (see the text). {\it Lower panel:} The difference between observed and synthesized  variations vs. time. 
}
\label{rv_time}
\end{figure*}

The best fit for the spot model with $Q=0.0$ is found for a RV offset of $21.6 \pm 1.0$~m~s$^{-1} $ and 
$\langle C_{\rm d} \rangle = 0.25 \pm 0.05 $. We plot the observed and simulated RV residuals vs. time in the upper panel of Fig.~\ref{rv_time} together with their differences in the lower panel. The error bars of the synthesized residuals take into account the effects of the photometric accuracy, the differential rotation, and the photometric residuals. The nominal accuracy of MOST photometry is $\sim 1.1 \times 10^{-4}$ in relative flux units that implies an error in the estimated spot area of the order of $\sim 2.5 \times 10^{-4}$ leading to a RV  error of $\sim 0.75$ m~s$^{-1}$ when Eq.~(1) in \citet{Desortetal07} is applied to compute the maximum deviation. The differential rotation changes the longitudes of the starspots during the time intervals considered for the photometric modelling. Since our modelling assumes a fixed pattern of spots during each interval of $7.625$ days, our synthesized RV variations have systematic errors produced by the migration of the active regions during those time intervals. We estimate a maximum deviation of $1.5-2.0$~m~s$^{-1}$ depending on the spot migration rate. Finally, the small active regions that evolve on timescales of $1-2$ days as revealed by the residuals of the photometric best fit (see Fig.~\ref{lc_bestfit} lower panel) also contribute to the uncertainty of our synthesized RV values. Since their estimated areas are of the order of $0.2-0.3$ percent of the stellar disc, the estimated fluctuations are of $\sim 1.8$~m~s$^{-1}$ using the model of Sect.~\ref{rv_from_model}. In conclusion, the overall standard deviation of our synthesized RV values is $\sim 2.8-3.0$~m~s$^{-1}$. 

The  agreement between our synthesized RV variations for $Q=0$ and the observations is generally good, although our model cannot reproduce the oscillations shown by the data on timescales as short as $1-2$ days because it assumes a fixed configuration of active regions that is changed with a timestep of $3.8$ days. 
{ Assuming the standard deviation of the RV measurements given by the SOPHIE reduction pipeline, we find a reduced $\chi^{2}=1.57$ for the best fit of the 23 simultaneous data points.} On the other hand, considering the observed RV fluctuations on timescales shorter than $1.0$ day, we derive a standard deviation of $5.55$~m~s$^{-1}$ giving a reduced $\chi^{2}=1.18$. 

Including a facular  contribution, i.e., considering models with $Q > 0$, leads to an increase of the reduced $\chi^{2}$ of the best fit. This happens because faculae produce an increase of the amplitude of the RV  variations in our model leading to values that are systematically below the observations  before HJD~2454308.0 and systematically above for epochs later than 2454310.0. The deviations are of about $10-15$~m~s$^{-1}$ for the spot model with $Q=5$, and are already significant for $Q=1-2$. We conclude that the observed RV fluctuations induced by stellar activity favour a spot model with a negligible facular contribution, supporting the result found in Sect.~\ref{model_param} for the facular-to-spotted area ratio considering only the photometric observations. 

\begin{figure}
\centering
\includegraphics[width=5cm,height=8cm,angle=90]{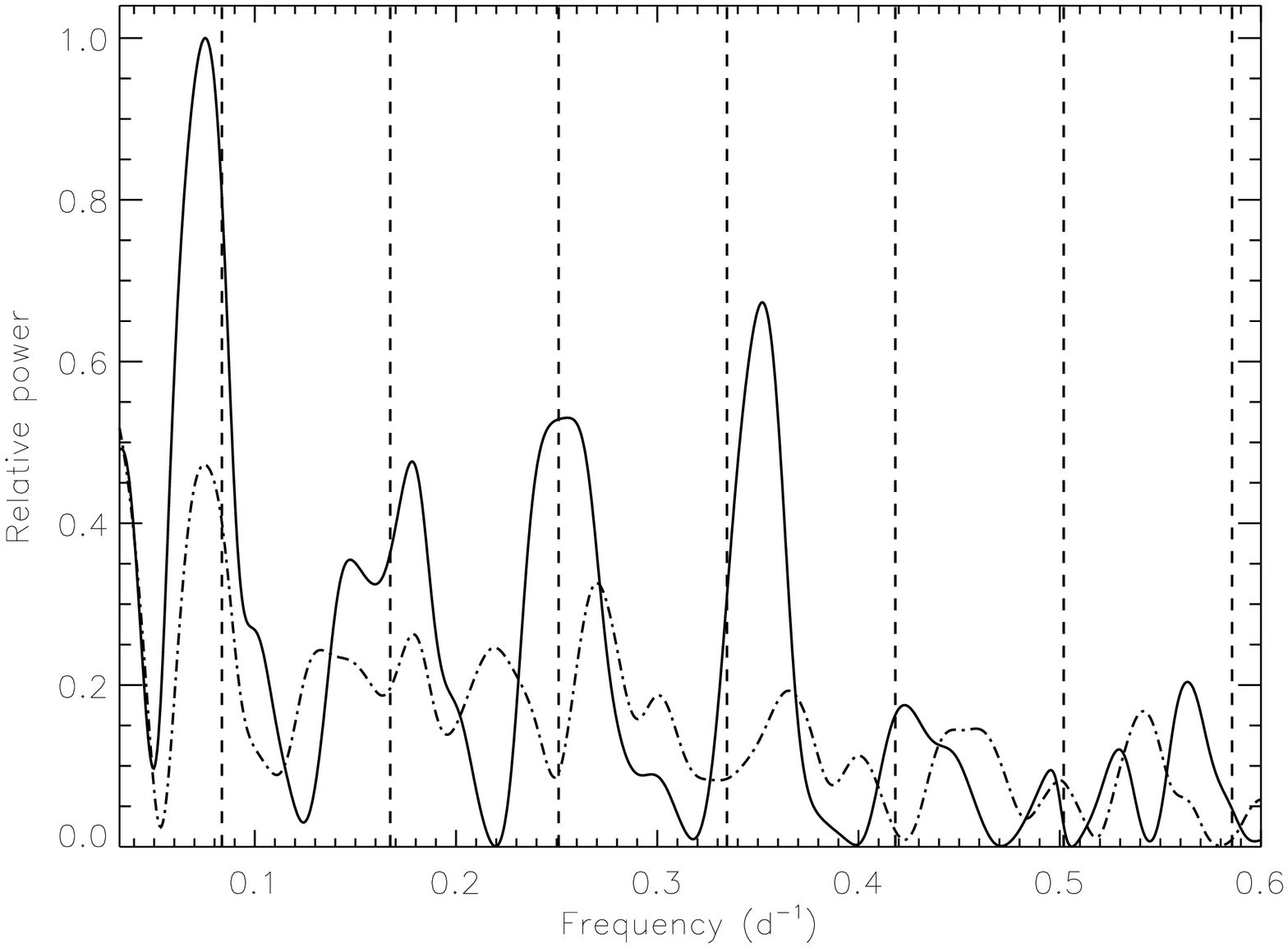} 
\includegraphics[width=5cm,height=8cm,angle=90]{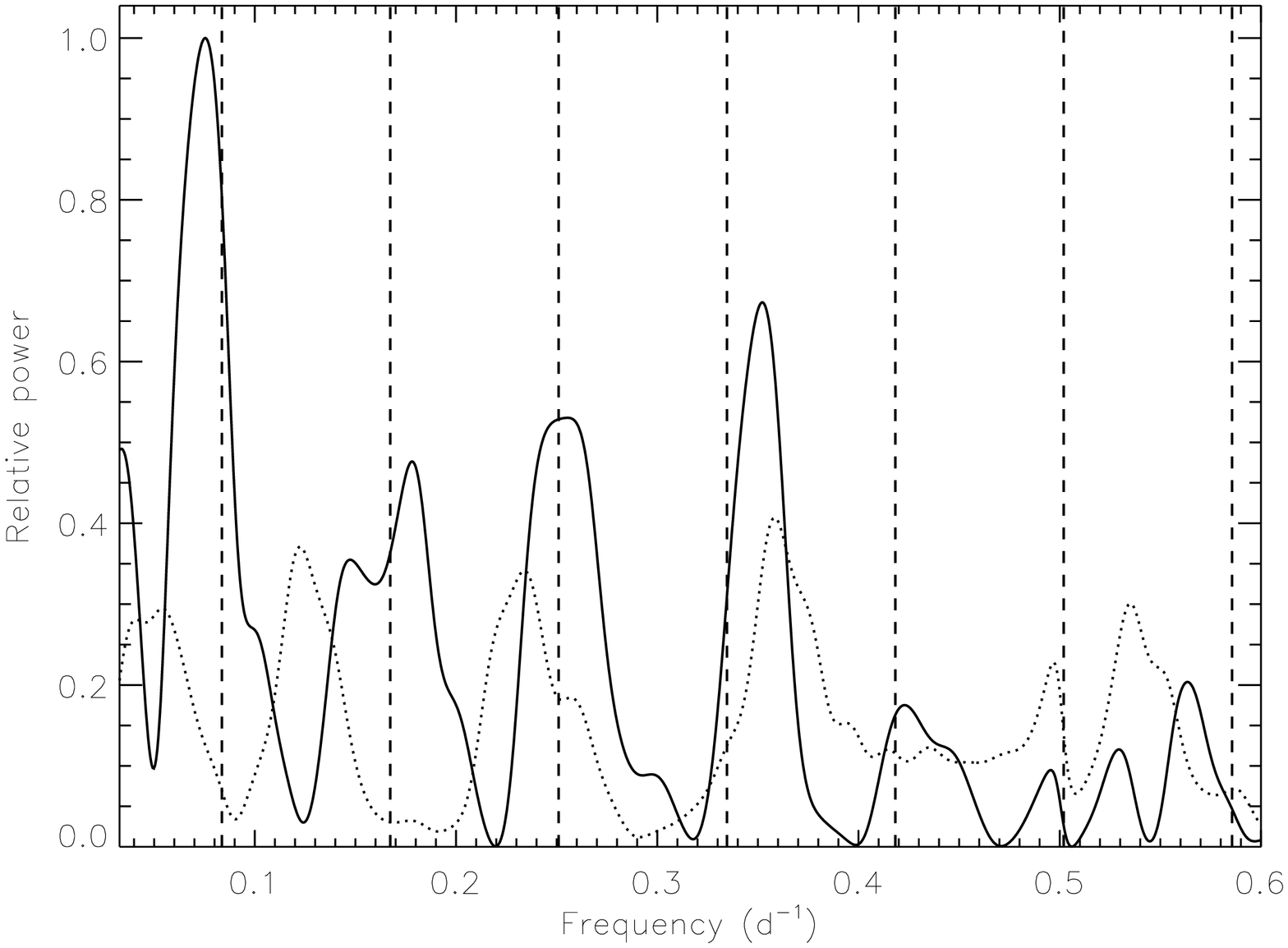} 
\includegraphics[width=5cm,height=8cm,angle=90]{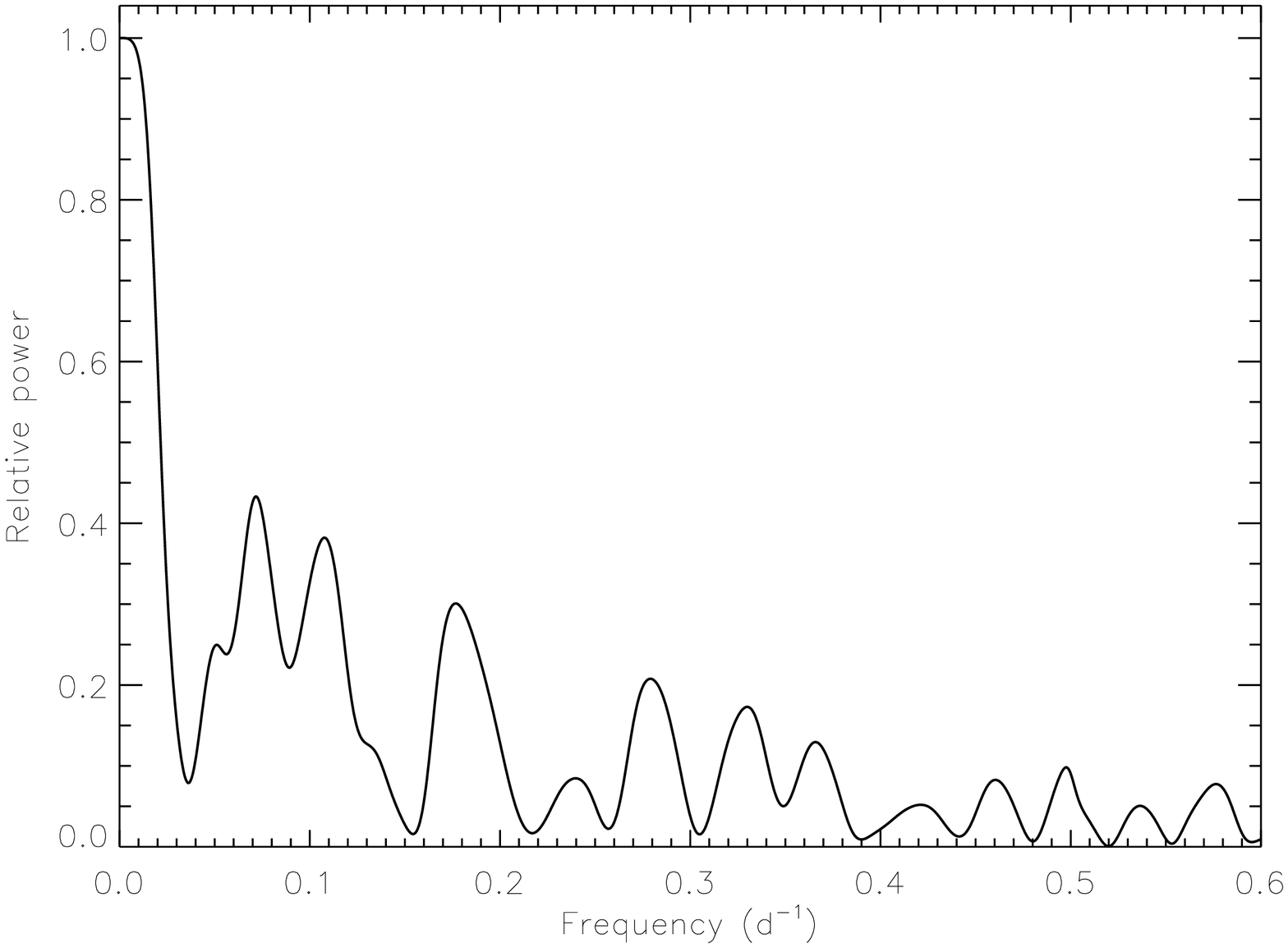} 
\caption{{\it Upper panel:} Lomb-Scargle periodogram of the observed RV variations produced by stellar activity  excluding the four data points between HJD~2454308.0 and 2454310.0 (solid line) and the periodogram of the corresponding synthetic time series (dash-dotted line); 
{\it Middle panel:}  Lomb-Scargle periodogram of the observed RV variations produced by stellar activity  excluding the four data points between HJD~2454308.0 and 2454310.0 (solid line) and the periodogram obstained after subtracting the synthesized RV variations from the observed RV time series (dotted line). The vertical dashed lines mark the rotation frequency and its harmonics for a rotation period of 11.953 days. {\it Lower panel:} The spectral window of the time series of the RV observations considered in the upper panel.
}
\label{power_spectrum}
\end{figure}

\citet{Boisseetal11} noticed that most of the power of the RV variations induced by stellar activity is concentrated at the rotation frequency and its first two or three harmonics. Although the number of data points is limited to 23 in our case, we applied their technique to test whether the subtraction of the synthesized RV  variations is capable of reducing the impact of stellar activity on the RV measurements.  We plot in Fig.~\ref{power_spectrum} the Lomb-Scargle periodogram of the time series of the RV measurements simultaneous with MOST photometry (solid line) and the periodograms of the  synthetic RV time series (dash-dotted line, upper panel) and of the residuals obtained after subtracting the synthetic series from the observations (dotted line, middle panel), excluding in all the cases the four points  between HJD~2454308.0 and 2454310.0. 
All the periodograms have the same normalization to allow an immediate comparison of their power levels. 
The  observed RV variations show the maximum power close to the rotation frequency with the successive harmonics having a slightly increasing power up to the third. This effect is likely due to the presence of multiple, partially unresolved peaks among which the power is split, as  clearly evident in the case of the first harmonic. Therefore, the peak of the third harmonic is higher than those of the first and the second because it is narrower.  The spectral window of the time series is plotted in the lower panel of Fig.~\ref{power_spectrum} and shows the presence of significant sidelobes, notably at $0.07$~d$^{-1}$ from the main peak. This, in combination with the remarkable differential rotation of HD~189733, can explain the multi-peaked features in the power spectrum, particularly evident in the signals of the first and second harmonics. 

When we subtract the synthesized RV variations,  the power is significantly reduced at the rotation frequency and its first and second harmonics, { although some residual power is still present between the fundamental frequency and the first harmonic and between the first and the second harmonics, probably as a consequence of the multiple frequency nature of those peaks in combination with the spectral window.} The reduction at the third harmonic after the subtraction of the synthetic RV time series is less remarkable. 
This is expected because our approach cannot model the variations on timescales shorter than $4-5$ days. Unfortunately, the small number of simultaneous RV and photometric observations limits our capability to test the performance of our approach. Nevertheless, the present application demonstrates that it can be highly advantageous even for a small sample of data with limited precision, i.e., a photometric precision of  a few percent of the flux modulation and a RV precision of $3-5$ m~s$^{-1}$.  


\section{Discussion and conclusions}
\label{conclusions}

We have applied the ME spot modelling method introduced by \citet{Lanzaetal09a} and \citet{Lanzaetal09b} to the MOST photometry of  HD~189733. It is among the brightest and most active stars accompanied by a transiting hot Jupiter and displays activity-induced RV variations  with an amplitude reaching  several tens of m~s$^{-1}$. The simultaneous RV measurements collected with SOPHIE offer us a unique opportunity to compare the models proposed to account for the RV variations induced by activity with the observations. 
Both the MOST photometry and the SOPHIE RV measurements have errors comparable with those expected with larger telescopes on fainter stars. In the case of CoRoT or Kepler, it is possible to reach a photometric relative accuracy of $1.5 \times 10^{-4}$ and $2.0 \times 10^{-5}$ on a G2V star of magnitude $V=12$ in one hour integration time, respectively. Using HARPS, we can improve the radial velocity accuracy down to $2-5$ m~s$^{-1}$ on stars of that apparent magnitude, making our test particularly relevant for those targets, especially for the most active ones, given that HD~189733 is among the most active planetary hosts found so far. 

Our spot models show several active regions with lifetimes comparable or longer than  the duration of  MOST observations, i.e., $\sim 30$ days, that evolve remarkably and migrate at different rates with respect to a reference frame rotating with the mean stellar rotation period. This is interpreted as evidence of surface differential rotation with a relative amplitude of $\Delta \Omega / \Omega = 0.23 \pm 0.10$, in agreement with the result obtained from Doppler Imaging \citep{Moutouetal07,Faresetal10}. It is interesting to note that, although the  MOST observations cover only $\sim 2.5$ stellar rotations, the differential rotation signal is fairly evident thanks  to its large amplitude. Moreover, a population of small active regions with a surface of $0.1-0.3$ percent of the stellar disc and typical lifetimes of $1-2$ days is observed, as indicated by the  oscillations in the residuals of the photometric model  (cf. Fig.~\ref{long_distr}). 

We introduce a method to synthesize the RV variations induced by stellar activity from the longitude distribution
of the stellar active regions derived by the ME spot modelling. The model assumes that the stellar active regions have the same contrasts and turbulent convection properties of the solar active regions. 
Starspots are assumed to be close to the stellar equator because we cannot derive  their latitudes from the high-precision photometry. Indeed, in the case of HD~189733  the spot latitudes could be estimated from the migration of the  active longitudes, given the differential rotation law of \citet{Faresetal10}, but the results are uncertain and we prefer to compute the RV variations adopting nearly equatorial spots to test our approach in the general case when the latitudinal dependence of the rotation period is unknown. 

Our model has only two free parameters, i.e., an offset  applied to match the zero points of the synthesized and observed RV scales, and a coefficient $\langle C_{\rm d} \rangle $ measuring the average reduction of the depth of the spectral lines in dark spots. They are derived by a best fit to the RV observations. 
The synthesized RV variations obtained from the starspot distributions  show a remarkably good agreement with the simultaneously observed RV variations, although the latter display remarkable changes on timescales as short as $1-2$ days that cannot be accounted for by our model whose spot distributions can be updated only with a time interval of $\sim 4$ days. This is a limitation of every model based on the rotational modulation of spot visibility because it requires a minimum time interval, not too short in comparison with the stellar rotation period, to map the brightness inhomogeneities on the stellar surface. 

The RV variations set also a constraint on the facular effects suggesting that dark spots are dominating both the photometric and the activity-induced RV variations, at least during the time interval covered by the present observations. Of  course,  this does not exclude the possible presence of bright faculae in the photosphere of HD~189733, but their contribution cannot be detected in the available data. We show that the best value of $Q$, the facular-to-spotted area ratio, is $Q=0$, with an acceptable range extending from~$\sim 0$~to~$\sim 5$ when only the modelling of the optical flux variations is considered. Considering the modelling of the RV variations, an upper limit of $Q \sim 2-3$ is derived.  In the case of the Sun, the best value of $Q$ is $9$ \citep{Lanzaetal07}.
Thus our results indicate a lower relative contribution of the faculae to the  light variation  of HD~189733. 
The amplitude of the rotational modulation of our star was $\sim 0.015$ mag during the present MOST observations and $\sim 0.03$ mag during the 2006 observations described in \citet{Miller-Riccietal08}, i.e., $\sim 5-10$ times that of the Sun at the maximum of the eleven-year cycle. This indicates that HD~189733 is remarkably more active than the Sun, which may account for the undetectable facular contribution to its light variations, as suggested by \citet{Radicketal98} and \citet{Lockwoodetal07}. 

We conclude that the approach presented in this paper can reduce the power of the activity-induced RV variations at the rotation frequency and its low-order harmonics by a factor ranging from 2 to 10, which represents an important improvement for the detection of exoplanets corotating with the star or with an orbital frequency in the $2:1$ ratio  with the rotation frequency, as suggested by the statistical studies of \citet{Lanza10} and \citet{Lanzaetal11}. Thanks to the high-precision space-borne photometry already available with MOST, CoRoT, and Kepler, it is possible to map the longitudinal distribution of active regions in late-type stars and apply the method presented in this paper to reduce remarkably the impact of stellar activity on their RV jitter  allowing us to confirm the detection of telluric planets or refine the measurement of their mass \citep[see, e.g., the application to CoRoT-7 in ][]{Lanzaetal10}.

\begin{acknowledgements}
{ The authors are very grateful to  the MOST team for making available the observations analysed in this work. They wish to thank the anonymous referee for valuable comments that helped to improve this work.}
Active star research and exoplanetary studies at INAF-Osservatorio Astrofisico di Catania and Dipartimento di Fisica e Astronomia dell'Universit\`a degli Studi di Catania 
 are funded by MIUR ({\it Ministero dell'Istruzione, dell'Universit\`a e della Ricerca}) and by {\it Regione Siciliana}, whose financial support is gratefully
acknowledged. 
This research has made use of  the ADS-CDS databases, operated at the CDS, Strasbourg, France.
IB would like to thank the support by the Funda\c{c}\~ao para a Ci\^encia e a Tecnologia (FCT), Portugal, through a Ci\^encia\,2007 contract funded by FCT/MCTES (Portugal) and POPH/FSE (EC), and in the form of grants reference PTDC/CTE-AST/098528/2008 and PTDC/CTE-AST/098604/2008. 
\end{acknowledgements}

\end{document}